\documentclass[%
 aip,numerical
 jmp,%
 amsmath,amssymb,
 reprint,%
]{revtex4-1}

\usepackage{graphicx}
\usepackage{dcolumn}
\usepackage{bm}
\usepackage{hyperref}

\begin{document}

\preprint{AIP/123-QED}

\title{Self-aligned graphene field-effect transistors with polyethyleneimine doped source/drain access regions}

\author{Hema C. P. Movva}
 \email{hemacp@utexas.edu}
\author{Michael E. Ram{\'{o}}n}
\author{Chris M. Corbet}
\author{Sushant Sonde}
\author{Sk. Fahad Chowdhury} 
\affiliation{ 
Microelectronics Research Center, The University of Texas at Austin, Austin, TX 78758
}%

\author{Gary Carpenter}
 \affiliation{
 IBM Research, Austin, TX 78758
}%

\author{Emanuel Tutuc}
\author{Sanjay K. Banerjee} 
\affiliation{ 
Microelectronics Research Center, The University of Texas at Austin, Austin, TX 78758
}%

\date{\today}

\begin{abstract}
We report a method of fabricating self-aligned, top-gated graphene field-effect transistors (GFETs) employing polyethyleneimine spin-on-doped source/drain access regions, resulting in a 2X reduction of access resistance and a 2.5X improvement in device electrical characteristics, over undoped devices. The GFETs on Si/SiO$_2$ substrates have high carrier mobilities of up to 6,300 cm$^2$/Vs. Self-aligned spin-on-doping is applicable to GFETs on arbitrary substrates, as demonstrated by a 3X enhancement in performance for GFETs on insulating quartz substrates, which are better suited for radio frequency applications. 
\end{abstract}

\maketitle

Graphene's exceptional electronic properties, in particular its high carrier mobility \cite{RiseGraph}, high saturation velocity \cite{ElecGraph} and large current density \cite{GraphCurrDen} make it an excellent channel material for future ultra-high-speed electronic devices. Graphene field-effect transistors (GFETs) have recently attracted significant attention for potential radio frequency (RF) applications, \cite{KoswattaRF} with reports of RF GFETs operating at  intrinsic cut-off frequencies up to 300 GHz \cite{Liao300}. However, theoretical calculations predict GFETs to be capable of running at THz operating frequencies \cite{LiaoSA}. One of the major factors preventing optimal THz RF performance of GFETs is high series resistance of the access regions between the source/drain electrodes and the top-gated graphene channel. This parasitic access resistance reduces device drive currents (I$_D$) and transconductances (g$_m$), thereby directly affecting GFET performance. Electrostatic modulation of the access regions is one way of reducing this resistance for devices on heavily doped substrates \cite{Jenkins50}. This approach requires a global, conducting back-gate and does not provide for independent control of multiple devices on the same substrate, and is unsuitable for GFETs on insulating substrates.  This problem can be addressed by fabricating GFETs with self-aligned gates, where the access region dimensions and consequently their resistances are reduced. Previous methods of self-aligned GFET fabrication are not straightforward, and either use physically assembled nanowire gates \cite{Liao300}, or require pristine quality graphene and ultra high vacuum metal-deposition \cite{FarmerSA}, or restrict scaling of the top-gate dielectric \cite{BadSA}. In this work, we present a simple and scalable approach of fabricating self-aligned GFETs, exploiting the unique property of charge-transfer doping of graphene using polyethyleneimine (PEI).

The problem of high access resistance in conventional Si complementary metal-oxide-semiconductor (CMOS) transistors is overcome by heavily doping the source/drain access regions through ion implantation \cite{BowerSA}. We employed a similar approach in this work, by using charge-transfer doping of graphene with PEI \cite{FarmerEH}, to dope the access regions and reduce their resistance.  We chose PEI due to its ease of application, but any other method of doping (substitutional, surface-transfer, \cite{HLiuDoping} etc.) can, in principle, be used for the same result. We fabricated top-gated GFETs with exposed source/drain access regions and doped them with PEI in a self-aligned manner using a controllable method of spin-on-doping. The fabrication process is universally applicable to GFETs on arbitrary substrates and is specifically demonstrated for GFETs on Si/SiO$_2$ and quartz substrates.

\begin{figure}
\centering
\includegraphics[scale=1.0,angle=270]{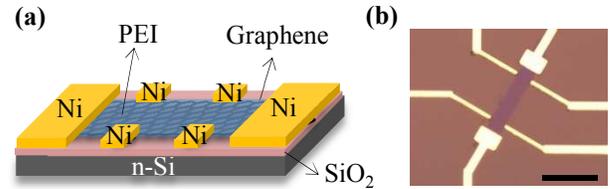}
\caption{(Color online) (a) Schematic of a PEI spin-on-doped back-gated GFET on Si/SiO$_2$ and (b) optical micrograph of a representative spin-on-doped GFET (scale bar is 10 $\mu$m)}
\label{fig:fig0}
\end{figure}

PEI spin-on-doping of graphene was characterized on back-gated GFETs. Monolayer graphene was exfoliated onto highly-doped n-type Si substrates with a 285 nm thermally grown SiO$_2$ layer acting as the back-gate dielectric. Graphene active regions were patterned by e-beam lithography (EBL) and oxygen plasma etching. A second EBL step was performed to define metal contacts for 4-point probe structures, followed by e-beam evaporation of a 50 nm Ni layer as contact metal and a final lift-off. The branched PEI (Sigma Aldrich, $M_{n} \sim$ 60,000, $M_{w} \sim$ 750,000) molecular dopants were applied by spin-coating a dilute solution of PEI in methanol onto the substrate. A solution of 0.02\%  (by wt.) PEI in methanol was prepared by magnetic stirring in a dark, air-tight container for a period of 48 hours. Methanol was used as a solvent due to its high volatility which minimizes solvent residue on the graphene surface. To dope the graphene, this dopant solution was spin-coated onto the GFETs at 1500 rpm for 60 s, followed by a quick bake at 90$^{\circ}$C for 20 s to drive away any remaining methanol residue. PEI, being a heavy macromolecule does not evaporate, but forms a thin adsorbed layer on the graphene, thereby doping it. Figure~\ref{fig:fig0} (a) shows the schematic and  Fig.~\ref{fig:fig0} (b) the optical micrograph of a back-gated GFET after spin-on-doping. PEI-doped graphene has a thin, uniform layer of dopants adsorbed on it and cannot be optically distinguished from clean, undoped graphene.

\begin{figure}
\centering
\includegraphics[scale=1.0]{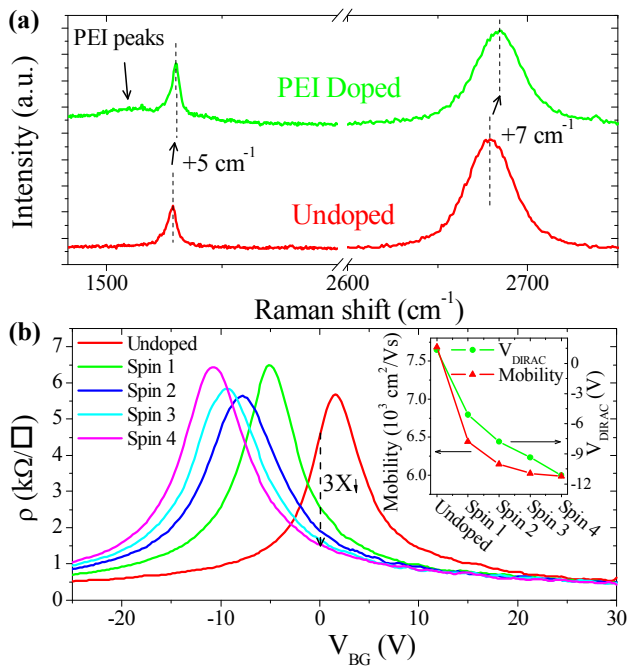}
\caption{(Color online) (a) Raman spectra of graphene before and after doping with PEI. There is a reduction in $I_{2D}$/$I_{G}$ from 2.5 to 1.5 and an upshift of the G and 2D peaks after doping. PEI characteristic peaks appear at $\sim$ 1500 cm$^{-1}$. (b) 4-point resistivity measurements of the GFET after repetitive spin-on-doping steps. There is a reduction in the extracted carrier mobility with successive spin-on-doping steps (inset).}
\label{fig:fig1}
\end{figure}

Figure~\ref{fig:fig1} (a) compares the Raman spectra of graphene before and after PEI-doping. The spectrum before doping is indicative of monolayer graphene with a 2D-to-G peak intensity ratio, $I_{2D}$/$I_{G}$ of 2.5 and a 2D band full width at half maximum (FWHM) of 25 cm$^{-1}$ \cite{RamanGupta}. Peaks around 1500 cm$^{-1}$ appear in the Raman spectrum after doping. These peaks indicate the presence of PEI and have been previously reported on carbon nanotube/PEI composite fibers \cite{PEIRaman}. Reduction in the $I_{2D}$/$I_{G}$ ratio to 1.5 and an upshift of both the G and 2D peaks is a further indication of n-doping\cite{RamanDoped,RamanElecDope}. The PEI molecules are weakly bonded to the graphene surface and are prone to slow desorption under ambient conditions \cite{FarmerEH}. However, the dopants can be sealed on the graphene surface using an atomic layer deposition (ALD) oxide capping layer to prevent desorption and ensure long-term stability of doping \cite{Capping}.

Control of the doping dose was achieved by repetitive spin-on-doping steps. Figure~\ref{fig:fig1} (b) shows 4-point resistivity measurements of the back-gated GFET after successive PEI spin-on-doping steps. The undoped GFET shows a Dirac point (V$_{DIRAC}$)  at + 3 V, likely due to unintentional doping by water vapor during the fabrication process \cite{GraphGeim}. V$_{DIRAC}$ shifts to - 5 V after the first spin-on-doping step, signifying n-type doping due to PEI. Successive spin-on-doping steps increasingly dope the graphene n-type, as evident from shifts in V$_{DIRAC}$ to larger negative voltages after every spin. Carrier mobilities were extracted using a well-established field-effect mobility model \cite{SKimAPL} and plotted along with V$_{DIRAC}$ after every spin-on-doping step in Fig.~\ref{fig:fig1} (b) (inset). There is a reduction in carrier mobility with each spin-on-doping step, in accordance with previous reports of dopant induced mobility degradation \cite{FarmerEH}. However, the mobility after four spin-on-doping steps still remains high at $\sim$ 6,000 cm$^2$/Vs. An important effect of doping is reduction in the graphene resistivity at V$_{BG}$ = 0 V from 4.5 k${\Omega}/{\square}$ to 1.5 k${\Omega}/{\square}$, a factor of 3X. This reduction in resistivity when employed to the source/drain access regions of a top-gated GFET can result in improved GFET performance.

\begin{figure*}
\includegraphics[scale=1.0,angle=270]{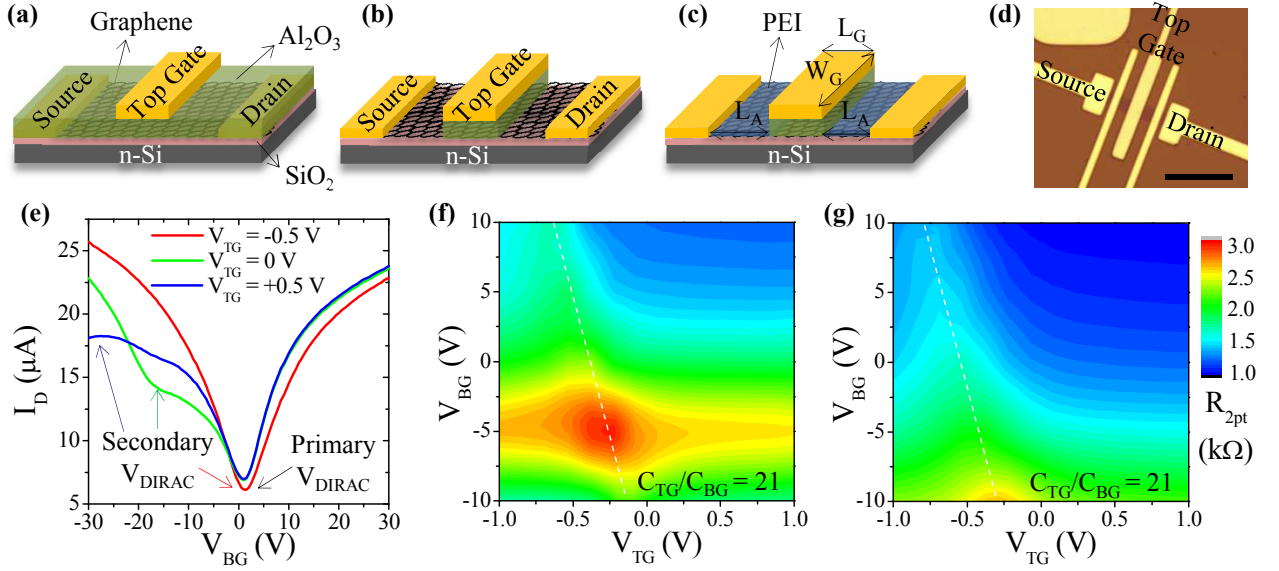}
\caption{(Color online) (a) Schematic of an as-fabricated top-gated GFET with ALD Al$_2$O$_3$ on top of the source/drain access regions and electrodes, which was (b) etched with 1:50 dilute HF in a self-aligned manner using the top-gate metal as a hard mask, and (c) subsequently spin-on-doped with PEI. The critical device dimensions are shown. (d) An optical micrograph of a representative spin-on-doped GFET (scale bar is 10 $\mu$m). (e) Drain current (I$_D$) profiles of a GFET prior to spin-on-doping, as a function of V$_{BG}$, at different V$_{TG}$, show selective modulation of the top-gated graphene region, independent of the access regions. (f) and (g) show the resistance contour plots for a GFET before self-aligned etching and after spin-on-doping respectively. The slope of the contours of charge neutrality (dashed lines) gives C$_{TG}$/C$_{BG}$ = 21.}
\label{fig:fig2}
\end{figure*}

To test this hypothesis, dual-gated GFETs were fabricated on clean, back-gated GFETs by patterning a top-gate stack. A 20 nm ALD Al$_2$O$_3$ layer, seeded by a 15$\AA$ evaporated Al layer, was deposited as the top-gate dielectric \cite{SKimAPL}. This was followed by a final EBL step to pattern the top-gate electrode. A 50 nm Ni layer was subsequently deposited as the top-gate metal contact. The source/drain access regions of the GFET at this stage were covered by the ALD Al$_2$O$_3$ layer, which was etched away in order to dope them. This etch was performed in a self-aligned manner using a 1:50 dilute HF solution as the etchant. The top-gate metal acts as a self-aligned hard mask during the etch and protects the top-gate dielectric. It has to be noted that the HF solution etches the underlying SiO$_2$ layer too, albeit at a much slower rate (5 nm/min.) than the ALD Al$_2$O$_3$ top-gate dielectric (60 nm/min.). The etch could thus be conveniently timed to minimize etching of the SiO$_2$ layer. The access regions were then finally spin-on-doped using PEI. Figures~\ref{fig:fig2} (a)-(c) illustrate the fabrication of self-aligned spin-on-doped GFETs and Fig.~\ref{fig:fig2} (d) shows an optical micrograph of the finished top-gated GFET. 

The drain current (I$_D$) profiles of a GFET after self-aligned etching, but prior to spin-on-doping, as a function of the back-gate bias (V$_{BG}$), at different top-gate biases (V$_{TG}$) are shown in Fig.~\ref{fig:fig2} (e). The profile at V$_{TG}$ = 0 V shows two V$_{DIRAC}$: a primary V$_{DIRAC}$ at 0 V and a secondary V$_{DIRAC}$ at $\sim$ -17 V. The primary V$_{DIRAC}$ arises from the source/drain access regions and the secondary V$_{DIRAC}$ from the top-gated graphene region. The n-type doping of the top-gated graphene is unintentional, and is probably due to impurities in the ALD Al$_2$O$_3$ layer. The position of the secondary V$_{DIRAC}$ can be selectively modulated using the top-gate bias. A negative V$_{TG}$ (= -0.5 V) depletes electrons from the top-gated graphene region and shifts the secondary V$_{DIRAC}$ close to 0 V, which, in turn, overlaps with the primary V$_{DIRAC}$ to result in one ``apparent" V$_{DIRAC}$ for the entire graphene. A positive V$_{TG}$ (= +0.5 V), on the other hand, induces excess electrons in the top-gated graphene region and shifts its V$_{DIRAC}$ to a more negative voltage, beyond - 30 V in this case, as evident from the downturn of I$_D$ around - 30 V. The primary V$_{DIRAC}$ is unaffected by the top-gate bias, signifying that the Dirac points of the top-gated graphene and the access regions can be modulated independent of each other.

Selective spin-on-doping of the GFETs was done only in the exposed source/drain regions using 0.02\% PEI, while the top-gated graphene region was protected by the top-gate stack. Figures~\ref{fig:fig2} (f) and (g) show the resistance contour plot of the device as a function of V$_{TG}$ and V$_{BG}$ before the self-aligned etch and after self-aligned doping. The only effect of self-aligned doping is to dope the access regions n-type and shift their V$_{DIRAC}$ to a negative voltage. This is evident as an apparent downward shift of the resistance profile along the V$_{BG}$ axis after doping. The resistance profile along the V$_{TG}$ axis remains the same before and after doping, signifying that there is no effect of doping on the top-gated graphene region. The dashed lines represent contours of charge neutrality for the top-gated channel and their slope gives the ratio of the top-gate capacitance to the back-gate capacitance, C$_{TG}$/C$_{BG}$ = 21. Using the back-gate capacitance value of C$_{BG}$ = 11 nF/cm$^2$, the top-gate capacitance is estimated to be C$_{TG}$ $\sim$ 236 nF/cm$^2$. This corresponds to a relative dielectric constant of 5.7 for the Al$_2$O$_3$ film, which agrees with reports from literature for ALD Al$_2$O$_3$ films on graphene \cite{BabakAPL}. The slope remains unchanged after the self-aligned etch and doping, thereby indicating no degradation of dielectric properties.

\begin{figure}
\centering
\includegraphics[scale=1.0,angle=270]{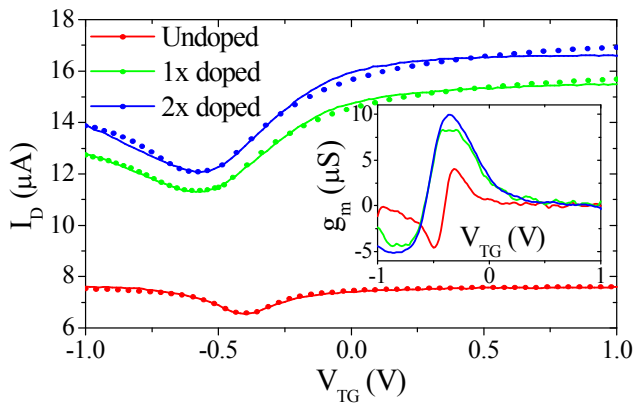}
\caption{(Color online) Transfer characteristics (at V$_{BG}$ = 0 V) of a top-gated GFET on Si/SiO$_2$ before and after one and two spins of 0.02\% PEI self-aligned doping show a 2.2X and 2.5X improvement in the maximum drive current (I$_{D,max}$) and peak transconductance (g$_{m,max}$) respectively (inset). The profiles are fitted to the mobility model described in Eq.~\ref{eq:Rprof} (circles) to extract a carrier mobility (${\mu}_{FE}$) $\sim$ 6,800 cm$^2$/Vs before doping and $\sim$ 6,300 cm$^2$/Vs after two doping spins.}
\label{fig:fig3}
\end{figure}

The properties of a GFET with a top-gated channel length, L$_{G}$ = 1.0 $\mu$m, channel width, W$_{G}$ = 7.0 $\mu$m and access region length, L$_{A}$ = 1.5 $\mu$m, before and after two self-aligned doping steps with 0.02\% PEI are shown in Fig.~\ref{fig:fig3}.  The transfer characteristics show an improvement in the maximum drive current (I$_{D,max}$) from 7.4 $\mu$A to 16.6 $\mu$A and in the peak transconductance (g$_{m,max}$) from 4.2 $\mu$S to 10.3 $\mu$S after two spins, an approximately 2.5X improvement. The I$_{ON}$/I$_{OFF}$ ratio also improves by 20\% after two spins. To extract the reduction in access resistance, the resistance profiles of the GFET ($R_{SD}$) are fit to the expression,

\begin{equation}
R_{SD} = R_S + \frac{L_G}{eW_{G}{\mu}_{FE}\sqrt{{n_0}^2 + n^2}}
\label{eq:Rprof}
\end{equation}

where $e$ is the electron charge, $n$ is the field-modulated carrier concentration, $n_0$ is the residual carrier concentration at the neutrality point, and ${\mu}_{FE}$ is the field-effect mobility of the top-gated graphene region \cite{SKimAPL}. $R_S$ is the total series resistance contribution from the source/drain contacts, $R_C$ and access regions, $R_A$; $R_S$ = $R_C$ + $R_A$. The undoped device shows a ${\mu}_{FE}$ $\sim$ 6,800 cm$^2$/Vs, which reduces to $\sim$ 6,300 cm$^2$/Vs after doping. $R_S$ reduces from 2.6 k$\Omega$ to 1.5 k$\Omega$ after one spin, and further down to 1.1 k$\Omega$ after two spins. This reduction in $R_S$ is primarily from reduction of $R_A$. It has to be noted that, in addition to improving the device characteristics, self-aligned spin-on-doping can be used as a knob to tune the I$_D$ of a GFET post-fabrication, by varying the doping dose. This method of tuning I$_D$ can be particularly useful to compensate for device-to-device variability in graphene integrated circuits \cite{Variability}.

\begin{figure}
\centering
\includegraphics[scale=1.0,angle=270]{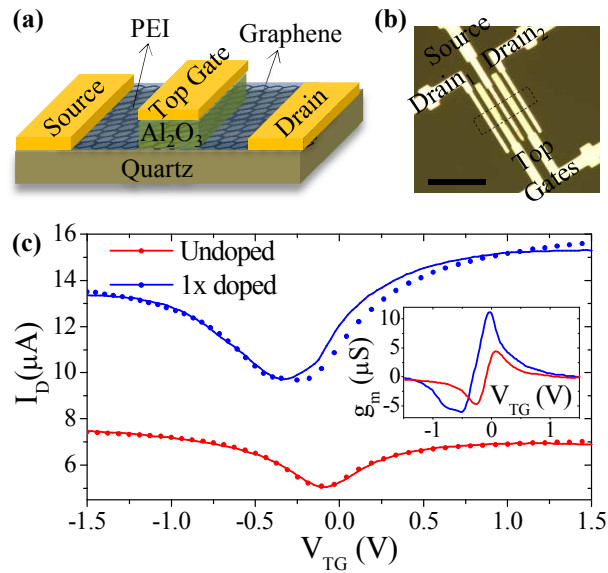}
\caption{(Color online) (a) Schematic of a self-aligned spin-on-doped GFET on quartz. (b) Optical micrograph of the GFET, with the graphene region marked by a dashed box. (scale bar is 10 $\mu$m) (c) The transfer characteristics show a $\sim$ 2.2X increase in I$_{D,max}$ and a $\sim$ 3X increase in $g_{m,max}$ (inset). The extracted carrier mobility (circles) reduces only slightly, from 5,600 cm$^2$/Vs to 5,200 cm$^2$/Vs after 0.02\% PEI doping.}
\label{fig:fig4}
\end{figure}

The full advantage of self-aligned spin-on-doping is realized on GFETs on insulating substrates, where the absence of a back-gate makes it impossible to electrostatically modulate the access region resistance\cite{SKimAPL}. Insulating substrates are also better suited for RF applications due to their lower parasitic capacitances. We specifically chose single crystal quartz substrates, since they are ideal for low loss, temperature stable high-frequency electronics \cite{MikeQz}. To fabricate GFETs on quartz, monolayer graphene was first exfoliated on Si/SiO$_2$ substrates and transferred onto quartz using a poly(methyl methacrylate) based transfer method \cite{KimTrans}. Subsequent processing was similar to the processing used for fabricating top-gated GFETs on Si/SiO$_2$ substrates.

Transfer characteristics of a top-gated GFET on quartz, before and after self-aligned doping using 0.02\% PEI are shown in Fig.~\ref{fig:fig4}. The device dimensions are: L$_{G}$ = 2.0 $\mu$m, W$_{G}$ = 7.0 $\mu$m and L$_{A}$ = 1.8 $\mu$m. The gains in device performance are similar to the GFET on Si/SiO$_2$. I$_{D,max}$ improves from 6.8 $\mu$A to 15.3 $\mu$A and g$_{m,max}$ from 4.3 $\mu$S to 13.1 $\mu$S, a $\sim$ 3 X improvement. A carrier mobility of 5,600 cm$^2$/Vs is extracted for the undoped device, which reduces slightly to 5,200 cm$^2$/Vs after doping. These values are comparable to the GFET on Si/SiO$_2$ and indicate that quartz can be an attractive insulating substrate for graphene electronics. 

In summary, we have demonstrated a simple and controllable method of spin-on-doping graphene using PEI as a chemical dopant. Control of the doping dose was achieved by repetitive spin-on-doping steps. We fabricated dual-gated GFETs on Si/SiO$_2$ substrates and spin-on-doped their access regions in a self-aligned manner to reduce their resistance and improve device performance by up to 2.5X. Further, we fabricated GFETs on insulating quartz substrates and spin-on-doped their access regions, which enhanced their device characteristics by up to 3X. These results indicate that chemical doping of the source/drain access regions can be a viable method of improving GFET performance on arbitrary substrates.

This work was supported by NRI-SWAN and the NSF NASCENT center. We thank D. Ferrer for help with TEM imaging. 

\nocite{*}


\providecommand{\noopsort}[1]{}\providecommand{\singleletter}[1]{#1}%

\end{document}